\documentclass[proceedings]{JHEP3}

\PrHEP{PrHEP hep2001}                   
\conference{International Europhysics Conference on HEP}                

\usepackage{epsfig}
\def\beq{\begin{equation}} 
\def\eeq{\end{equation}}
\def\berr{\begin{eqnarray}}
\def\eerr{\end{eqnarray}}

\title{Antimatter as a probe for physical processes in the early Universe}

\author{\speaker
{Alexander Sakharov}, Maxim Khlopov$^{\dag\ \ddag}$  and Sergei
Rubin$^{\dag\ \ddag}$\\
            
$^*$Swiss Institute of Technology, ETH--Z\"urich,  CH-8093 Z\"urich, Switzerland \\   
$^{\dag}$ Moscow Engineering Physics Institute, Kashirskoe shosse 31, 115409
Moscow,  Russia \\
$^{\ddag}$ Center for CosmoParticle Physics "Cosmion", Miussk. pl. 4, 125047
Moscow, Russia\\           
    E-mail: \email{sakhas@particle.phys.ethz.ch},
\email{mkhlopov@orc.ru}, \email{serg.rubin@mtu-net.ru}}

\abstract{The whole set of astrophysical data indicates that our
Universe is globally baryon asymmetrical. Nevertheless
 a possibility of existence of relatively small amount of sufficiently large
antimatter regions is not excluded. Such regions can
survive the annihilation with surrounding matter only in the case if their
sizes exceed a certain scale. It is shown that quantum fluctuations
of a complex scalar field caused by inflation can 
generate  large antimatter domains progenitors, which contribute
insignificantly to the total volume of the Universe. The resulting
distribution and evolution of such antimatter regions could cause every
galaxy to be a harbour of an  anti--star globular cluster.  
The existence of one of such anti -- star globular cluster in our Galaxy,
does not contradict the observed $\gamma$ -- ray background, but the expected
fluxes of $\overline{^4He}$ and $\overline{^3He}$ from such an antimatter
object can be searched for in PAMELA experiment and  are definitely accessible for the sensitivity of coming AMS02
experiment.}

\begin{document}

\vspace*{-0.3cm}\section{Introduction}
It is now widely accepted that the Universe is baryon asymmetrical as a whole.
Indeed, if the Universe would contain equal amount of matter and antimatter domains, coexisting with each other, the annihilation on the border of these domains will not disturb the observed diffused $\gamma$--ray background only in the case when the characteristic size of domains exceeds $10^3$Mpc \cite{exl,kolb,crg}. This scale is comparable with the modern cosmological horizon, what suggests that baryon asymmetry is global over the whole volume of the Universe. 

  However, the above mentioned arguments cannot exclude the Universe composed almost entirely of matter with relatively small insertions of antimatter regions. If some antimatter regions are sufficiently big, they can survive until now and evolve into astrophysical antimatter objects \cite{khl}. 
Only domains with the present physical sizes exceeding the critical surviving size $L_c=8h^2$kpc \cite{we} survive the annihilation with surrounding matter and can evolve into astrophysical objects. 

In this report we consider the model of inhomogeneous baryogenesis \cite{zil} based on the inflationary evolution of the baryon charged scalar field, what makes reasonable to discuss the existence of an anti--star globular cluster in our Galaxy and its observational signature in AMS02 detector, which will be installed on the International Space Station \cite{ams}.

\section{Scenario of inhomogeneous baryogenesis with antimatter generation}

\FIGURE[br]{
  \epsfig{file=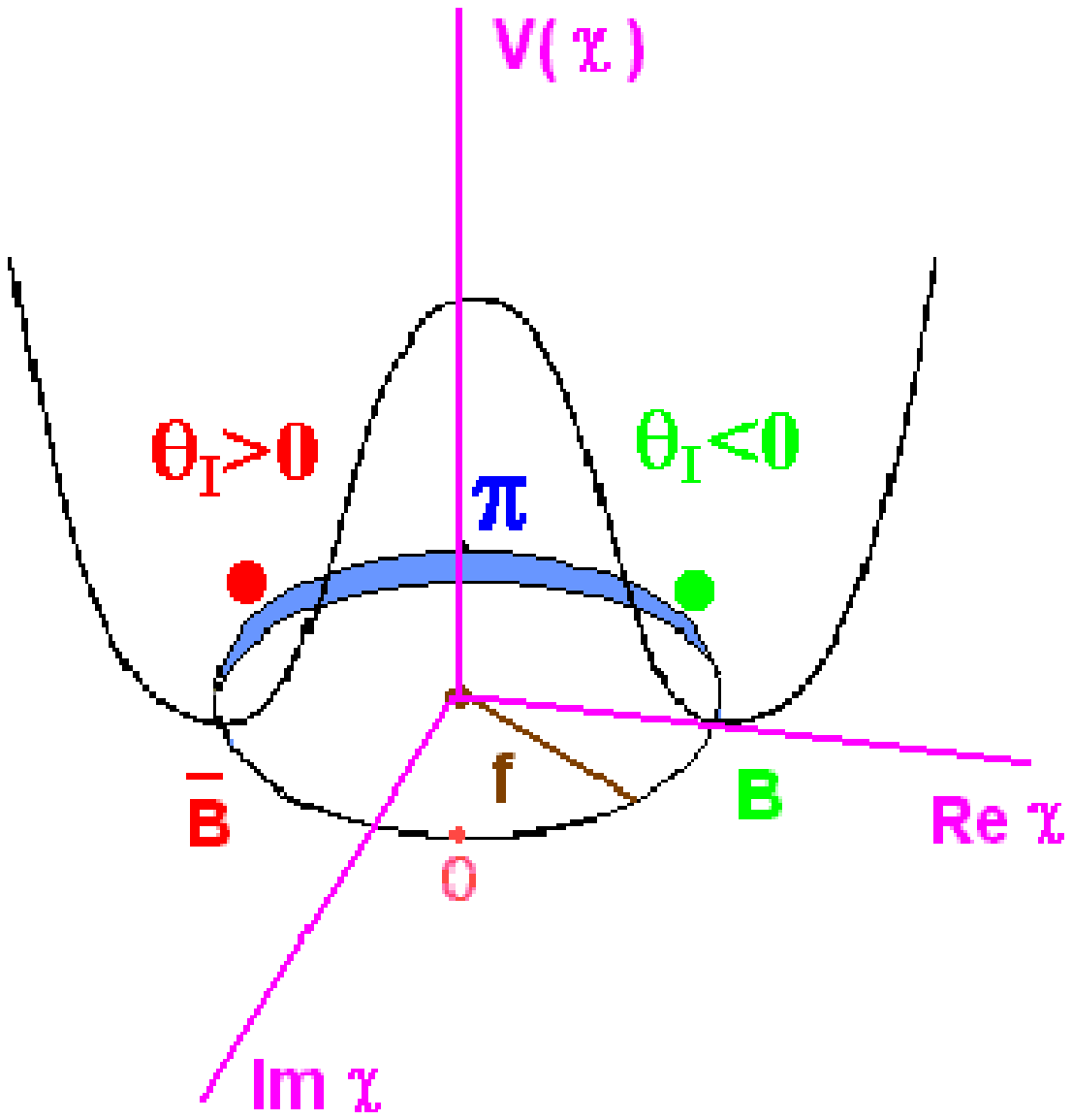,width=50mm} 
    \caption{PNG potential in the spontaneous baryogenesis mechanism. The sign of produced baryon asymmetry depends on the starting point of oscillations.\\ }
    \label{spont}}

\noindent Our approach \cite{zil} is  based on the
spontaneous baryogenesis mechanism \cite{sb}, which implies the existence
of a complex scalar field $\chi =(f/\sqrt{2})\exp{(\theta )}$ carrying
the baryonic charge. The $U(1)$ symmetry, which corresponds to the baryon charge, is broken spontaneously and explicitly. The explicit breakdown of $U(1)$ symmetry is caused by the phase
\noindent dependent term
\beq\label{expl}
V(\theta )=\Lambda^4(1-\cos\theta ),
\eeq
which results in the pseudo Nambu--Goldstone (PNG) potential Fig.\ref{spont}. The possible lepton number violating interaction of the field $\chi$ with matter fields can have the following structure \cite{dolgmain,zil}
\beq\label{leptnumb}
{\cal L}=g\chi\bar QL+{\rm h.c.}, 
\eeq
where fields $Q$ and $L$ represent a heavy quark and lepton, coupled to the ordinary matter fields. At the certain moment, in the early Universe, when the friction term, induced by the Hubble constant, becomes comparable with the angular mass $m_{\theta}=\frac{\Lambda^2}{f}$, the phase $\theta$ starts to oscillate around the minima of PNG potential and decay into matter fields due to (\ref{leptnumb}). The coupling (\ref{leptnumb}) gives rise to the following \cite{dolgmain,zil}: as the phase starts to roll down in the clockwise direction during the first oscillation Fig.\ref{spont}  it preferentially creates baryons over antibaryons, while the opposite is true as it starts to roll down in the opposite direction. The baryon/antibaryon number, created in these oscillations, is given by \cite{zil}
\beq\label{baryonnumber}
N_{B(\bar B)}\approx\frac{g^2f^2m_{\theta}}{8\pi^2}\Omega_{\theta_i}\theta_i^2\int\limits_{\mp\theta_i/2}^{\infty}d\omega\frac{\sin^2\omega}{\omega^2},
\eeq
where $\Omega_{\theta_i}$ is the volume, in which the phase has the value $\theta_i$. Thus, the distribution of the resulting baryon charge reflects the primordial distribution of the phase $\theta$ in the early Universe. 

We suppose \cite{zil} that the radial mass $m_{\chi}$ of the field $\chi$ is
larger than the Hubble constant $H_{infl}$ during inflation, while for the
angular mass of $\chi$ just the opposite condition $m_{\theta}\ll H_{infl}$ is
satisfied in that period. It provides the $U(1)$ symmetry to be
already broken spontaneously at the beginning of inflation, whereas the
background vacuum energy is still so high, that the cosmological friction term $3H_{infl}\dot\theta$ does not allow the phase to move classically, making the behaviour of $\theta$ akin to the behaviour of a massless scalar field.

\FIGURE[hr]{
  \epsfig{file=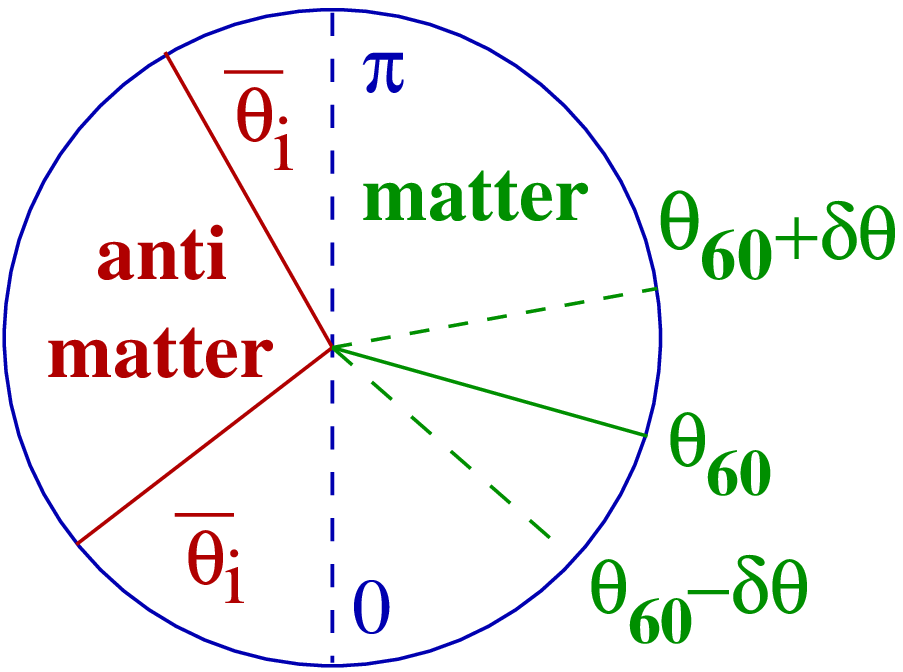,width=50mm} 
    \vspace*{-0.4cm}\caption{The phase makes Brownian step $\delta\theta 
=H/(2\pi f)$  at each e--fold. The phase is set to point $\theta_{60}$ in the range $[\pi
,0 ]$ at the beginning of inflation.  
The typical wavelength of the 
fluctuation $\delta\theta$   generated at this timescale is equal to
$H^{-1}_{infl}$, the whole domain  $H^{-1}_{infl}$,   containing $\theta_{60}$, after
one e--fold becomes divided into  $e^3$ causally
disconnected domains of radius $H^{-1}_{infl}$. Each new domain contains almost
homogeneous  phase value  $\theta_{60-1}=\theta_{60}\pm\delta\theta$. In
half of these  domains the phase evolves towards $\pi$
and in the other  domains  it moves  towards zero. Every successive e-fold this process repeats in every domain.}
    \label{fluct}} 

 \noindent This means that quantum fluctuations of $\theta$ at the de
Sitter background \cite{linde} will define the primordial phase distribution in the early Universe. In our case, such fluctuations can be interpreted as one--dimensional Brownian motion \cite{linde} along the circle
valley of PNG potential\footnote{The radius $f$ of PNG potential is equal to the scale of spontaneous U(1) symmetry breaking.} Fig.\ref{fluct}. Thus to have
the globally baryon dominated Universe the  phase should have the value in the
range $[\pi ,0 ]$ Fig.\ref{fluct}, just at the beginning of inflation\footnote{We start to consider inflation at 60 e--folds}  (when the size of the modern Universe crosses the horizon). Then subsequent quantum
fluctuations  move the phase to the values $\bar\theta_i$ in the range
$[0,\pi ]$ Fig.\ref{fluct} causing the successive antibaryon excess production. If it takes place not
later than after 15 e -- folds from the beginning of inflation
\cite{zil}, the physical contemporary size of antimatter domain progenitors will exceed the critical surviving size
$L_c=8h^2$kpc. The numerical simulation \cite{zil} of phase distribution in the inflationary Universe, at the condition $f\ge H\simeq 10^{13}$GeV,  shows, that a volume box corresponding to each galaxy can contain up to 10 above--critical progenitors with the phase $\bar\theta_i$ Fig.\ref{fluct}, what makes every galaxy a harbour of an antimatter domain. At the same time the fraction of
the total volume of the Universe containing $\bar\theta_i$ is only $\simeq 10^{-9}$ \cite{zil}, what makes sure that the Universe will
become baryon asymmetric as a whole. During the Friedman epoch the condition $m_{\theta}\ll H$ eventually gets
violated and the oscillations of $\theta$ are started, converting  the stored energy density $\rho_{\theta}\simeq
V(\theta_I)$ Fig.\ref{spont} into baryons
and antibaryons via coupling (\ref{leptnumb}). 
All those domains where the phase starts to oscillate from the
values $\bar\theta_i$ will contain antimatter. The antimatter density inside a domain depends on the initial value $\bar\theta_i$ Fig.\ref{fluct} and can be different in the
different domains \cite{zil}. The average number density of surrounding
matter should be normalised on the observable one $n_B/s\simeq 3\cdot
10^{-10}$. This normalisation sets the condition $f/m_{\theta}\ge 10^{10}$
for the PNG potential \cite{zil}.

\section{Evolution and observational signature of antimatter domains. Antimatter signals from dark matter particles.}
\vspace*{-0.2cm} 
The antibaryon number (\ref{baryonnumber}) in progenitors shows a strongly rising dependence on the initial value of the phase $\bar\theta_i$, what  makes sense to discuss the possibility to have a high density antimatter region in every galaxy. Let us consider the evolution of such high density antimatter region in the matter surrounding. 
\FIGURE[tr]{
  \epsfig{file=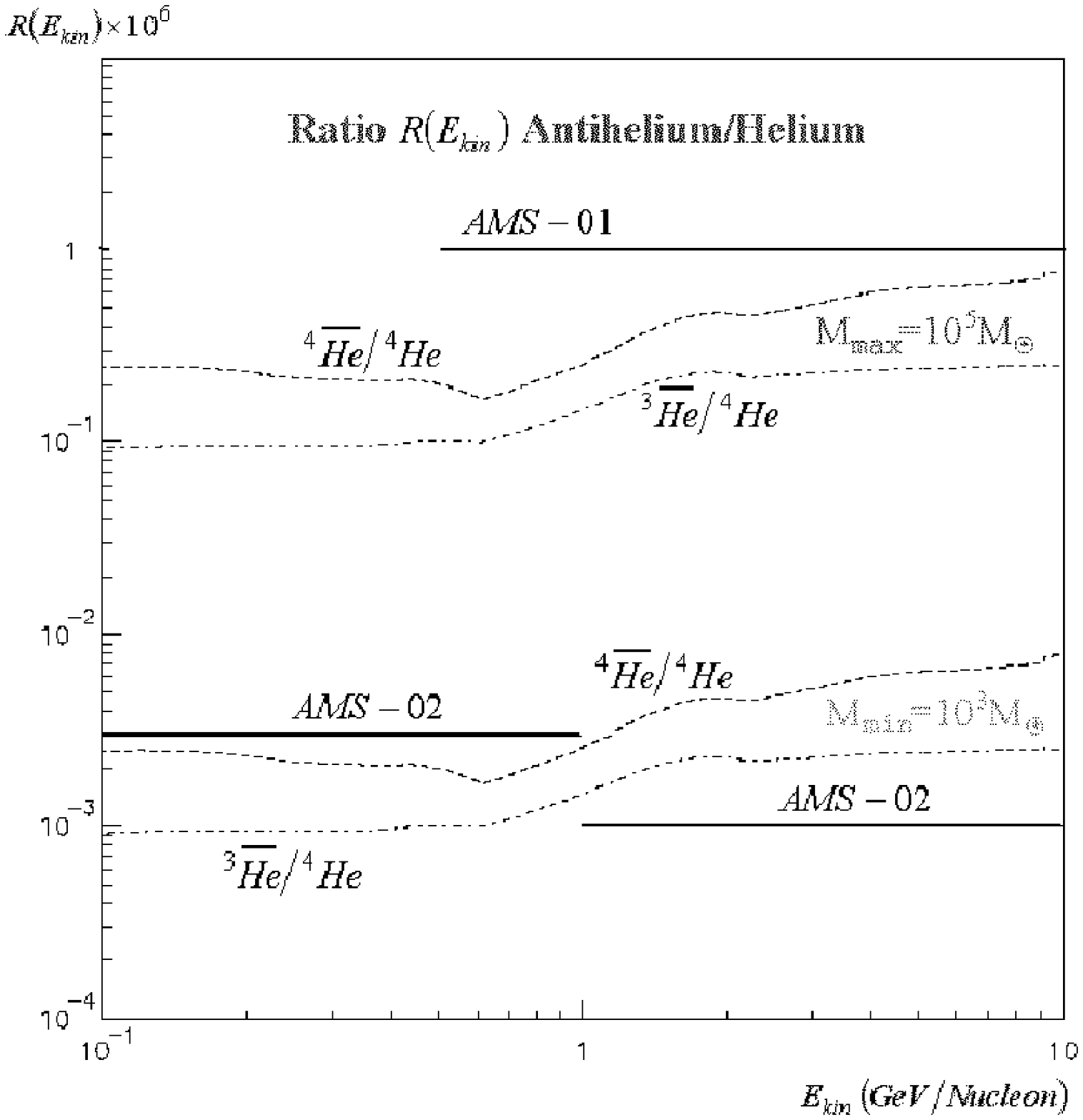,width=80mm} 
    \caption{The expected fluxes of  $\overline{^4He}$ and $\overline{^3He}$ from anti--star GC in the mass range  $10^3M_{\odot}-10^5M_{\odot}$ in the comparison with AMS02 sensitivity.}
    \label{antiheliumsignal}}
It is well known \cite{glob} that a cloud with the mass $10^5M_{\odot}-10^6M_{\odot}$, which has
the temperature near $10^4$K and the density several ten times that of the
 surrounding hot gas, are gravitationally unstable. This object is identified as a
protoobject of globular cluster (GC) and reflects the Jeans mass at the
recombination epoch. Thus if the primordial
antibaryon density inside the antimatter region progenitor was one order of magnitude higher then the 
surrounding matter density, that region can evolve into an antimatter GC
\cite{khl}. GC's
are the oldest galactic star systems to form in the Universe, and contain stars
of the first population. Thereby GC at the large galactocentric distance is the ideal
astrophysical objects which could be made out of antimatter.  The $\bar p$ releasing from such an anti-star GC by the stellar wind and anti--supernova explosions will be collected in our Galaxy and annihilate with $p$ contributing into GeV range diffused $\gamma$-- ray background \cite{khlopgolubkov}. This contribution being compared with the measured by EGRET $\gamma$-- ray background sets the upper limit on the mass of anti--star GC in our galaxy to $10^5M_{\odot}$ \cite{khlopgolubkov}, while $L_c$ defines the lower mass limit $10^3M_{\odot}$ on a possible anti-star GC. 

The most important experimental signature of the existence of an anti-star GC in our Galaxy, would be the observation of antinuclei in the cosmic rays near Earth orbit \cite{bgk}. The expected
fluxes of $\overline{^4He}$ and $\overline{^3He}$ Fig.\ref{antiheliumsignal} from such an antimatter
 object \cite{bgk} are only a factor two below the limit of AMS--01 (STS--91) 
experiment \cite{amsl} and definitely accessible for the sensitivity of
coming AMS--02 experiment. Thus one can conclude that AMS02 \cite{ams} experiment provides the test of nontrivial physical processes , underlying inflation and baryosynthesis. 

The annihilation in our galaxy of lightest supersymmetric particles, which can play the role of the cold dark matter, exposed as antimatter  fluxes (positrons and antiprotons)  at the earth \cite{lsp}. Space based PAMELA experiment \cite{pamela} provides high statistic measurements of these fluxes Fig.\ref{pamella}.  This experiment is accessible to search for antinuclei from an anti-star GC
and to positron fluxes from annihilation of sparse component of
stable 50 GeV neutrino of 4th generation, predicted by the superstring
phenomenology \cite{fargion}, especially, if such annihilation in the Galaxy is enhanced
by new Coulomb-like interaction, ascribed to the 4th generation \cite{4N}.  
\FIGURE[h]{
  \epsfig{file=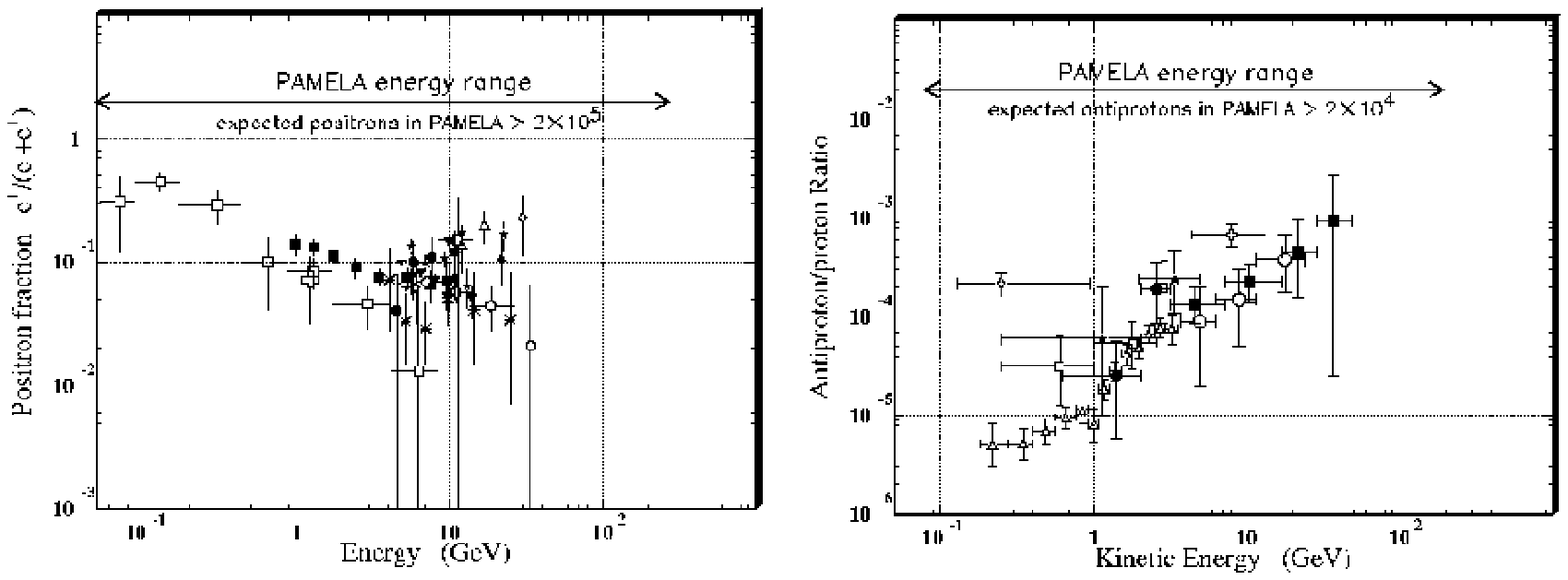,width=150mm} 
   \vspace*{-0.6cm}\caption{The ratios $p/p$ and $e^+/(e^+ +e^-)$ obtained with balloon--borne experiment. Up to now only about 500 $\bar p$ have been detected. PAMELA can provide high statistics measurements of both ratios, spanning over three decades in energy with the same detector.} 
    \label{pamella}}

\bigskip 
\noindent{\bf\underline{Acknowledgement}} The speaker (AS) acknowledges 
\underline{Mark Pears}, who kindly provided him with the expected PAMELA statistics. The work of (MKh) and (SR) was supported by project "Cosmoparticle
physics",
collaborations Cosmion--ETHZ, AMS--Epcos and kind hospitality (MKh) of
IHES.

\end{document}